\documentclass[showpacs,preprint,preprintnumbers,amsmath,amssymb]{revtex4}
\begin{document}

\title{Critical densities for the Skyrme type effective interactions}

\author{B. K. Agrawal,  S. Shlomo,  and V. Kim Au}
\affiliation{Cyclotron Institute, Texas A\&M University,
College Station, TX 77843, USA}
\begin{abstract}
We use the  stability conditions of the Landau parameters for the
symmetric nuclear matter and pure neutron matter to calculate the
critical densities for the Skyrme type effective nucleon-nucleon
interactions.  We find that the critical density can be maximized by
adjusting appropriately  the values of the  enhancement factor $\kappa$
associated with  isovector giant dipole resonance, the quantity $L$
which is directly related to the slope of the  symmetry energy and the
Landau parameter $G_0^\prime$.  However, restricting $\kappa$, $L$ and
$G_0^\prime$ to vary within acceptable limits reduces the maximum value
for the critical density $\tilde\rho_{cr}$  by $\sim 25\%$.  We also show
that among the various quantities characterizing the symmetric nuclear
matter, $\tilde\rho_{cr}$ depends strongly on the isoscalar effective
mass $m^*/m$ and surface energy coefficient $E_s$. For realistic values
of  $m^*/m$ and $E_s$ we get $\tilde\rho_{cr} = 2\rho_0$ to $ 3\rho_0$
($\rho_0 = 0.16$fm$^{-3}$).
\end{abstract} 
\maketitle


There are many different parameterizations of the Skyrme type effective
nucleon-nucleon  interaction as recently reviewed in Ref. \cite{Bender03}.
Values of the Skyrme parameters are usually  obtained by fitting the
results of the   Hartree-Fock calculations for the binding energies,
charge radii and spin-orbit splittings of a few closed shell nuclei
to the observed ones using a least square procedure.  The
Skyrme parameters are also constrained to yield appropriate values
for some properties of the infinite symmetric nuclear matter at the
saturation density $\rho_{nm}$. Over time, several improvements were
made by appropriately treating the center of mass correction to the
binding energy,  modifying the spin-orbit part of the interaction 
and including in the least square fit  the equation of state
(EOS) for the pure neutron matter obtained from a realistic interaction
\cite{Chabanat97,Chabanat98}. Nevertheless, not all the Skyrme
parameters are well determined in this way.

Recently, it has been shown in Ref. \cite{Margueron02} that  a more
realistic  parameterization of the Skyrme interaction can be obtained
by subjecting it to stability  requirements of the EOS defined by
the inequality conditions for the Landau parameters for symmetric
nuclear matter and pure neutron matter.  In other words, few of the
Skyrme parameters  not well determined by the experimental data used
in the least square procedure can be restricted by requiring that the
inequality conditions  are  satisfied  up to a maximum value for a nuclear
matter density,  also  referred to as the critical density $\rho_{cr}$.
A very recent systematic  study carried out in Ref.
\cite{Stone03} using several Skyrme interactions indicates that the
density dependence of the symmetry energy coefficient $J$ plays a
critical role in determining the properties of neutron star. Out of 87
different parameterizations for the Skyrme interaction considered in
Ref. \cite{Stone03}  only 27 of them, having a positive slope for the
symmetry energy coefficient at nuclear matter densities $\rho$ up to
$3\rho_0$ ($\rho_0 = 0.16$fm$^{-3}$),  are found to be suitable for the
neutron star model. Thus, it appears that the  parameters of  the Skyrme
interactions can be better constrained by combining the findings of Refs.
\cite{Margueron02} and \cite{Stone03}.

In the present work we study the dependence of $\rho_{cr}$  on the
nuclear matter saturation density $\rho_{nm}$, binding energy per nucleon
$B/A$, isoscalar effective mass $m^*/m$, incompressibility coefficient
$K_{nm}$, surface energy $E_s$, and symmetry energy coefficient $J$,
as also considered in Ref. \cite{Margueron02}. It must be
emphasized that we determine $\rho_{cr}$  in terms of the enhancement
factor $\kappa$, the coefficient $L = 3\rho dJ/d\rho$ and the Landau
parameter $G_0^\prime$ (at $\rho_{nm}$) instead of the combinations
$t_ix_i$ ($i=1, 2, $ and 3) of the Skyrme parameters $t_i$ and $x_i$ as
used in Ref. \cite{Margueron02}. Unlike the combinations $t_ix_i$, the
quantities $\kappa$, $L$ and $G_0^\prime$, which can be expressed in terms
of the Skyrme parameters,  are related to some physical processes. The
enhancement factor $\kappa$  accounts for the deviations from the
Thomas-Reiche-Kuhn (TRK) sum rule in the case of the isovector giant
dipole resonance.  The value of $\kappa $  at the $\rho_{nm}$ is expected
to be $\sim 0.5$ \cite{Chabanat97,Keh91}. The slope of the symmetry
energy coefficient must be positive for $\rho $ up to $ 3\rho_0$, so
that resulting Skyrme interaction can  be suitable for the study of
the properties of  neutron stars \cite{Stone03}. The Landau parameter
$G_0^\prime$ must be positive at  $\rho \le  \rho_{nm}$ in order  to
appropriately describe the position of the isovector $M1$ and Gamow-Teller
states \cite{Keh91}.  We find that by restricting the values of 
$\kappa$, $L$ and $G_0^\prime$ within the acceptable limits, the maximum
critical densities are lowered by about $25\%$ compared to the ones
obtained without such  restrictions \cite{Margueron02}.  The constrains
proposed in the present  work not only maximizes the value of the critical
density, but also ensures that the resulting Skyrme interaction can be
used to study the bulk properties of finite nuclei as well as those  of
neutron stars.

Recently \cite{Bender03,Bender02} a generalized form for the Skyrme
energy density functional has been obtained using the Hohenberg-Kohn-Sham
approach. However, in the present work we restrict ourselves to 
the energy density functional associated with the  commonly used Skyrme  
interaction \cite{Chabanat97,Vautherin72},
\begin{eqnarray}
\label{v12}
&V_{12}&= t_0\left (1+x_0 P_{12}^\sigma\right )\delta({\bf r}_1-{\bf
r}_2)\nonumber\\
&&+\frac{1}{2}t_1\left (1+x_1 P_{12}^\sigma\right ) 
\times \left[\overleftarrow{k}_{12}^2\delta({\bf r}_1-{\bf r}_2)+\delta({\bf
r}_1-{\bf r}_2)\overrightarrow{k}_{12}^2\right] \nonumber\\
&&+t_2\left (1+x_2 P_{12}^\sigma\right )\overleftarrow{k}_{12}\delta({\bf r}_1-{\bf r}_2)\overrightarrow{k}_{12} \nonumber\\
&&+ \frac{1}{6}t_3\left (1+x_3 P_{12}^\sigma\right )\rho^\alpha\left(\frac{{\bf r}_1+{\bf r}_2}{2}\right )
\delta({\bf r}_1-{\bf r}_2) \nonumber\\
&&+iW_0\overleftarrow{k}_{12} \delta({\bf r}_1-{\bf
r}_2)(\overrightarrow{\sigma_1}+
\overrightarrow{\sigma_2})\times \overrightarrow{k}_{12}
\end{eqnarray}
where, $P_{12}^\sigma$ is the spin exchange operator, $\overrightarrow{\sigma}_i$ is the Pauli spin operator, $\overrightarrow{k}_{12} = -i(
\overrightarrow{\nabla}_1-\overrightarrow{\nabla}_2)/2$
and $\overleftarrow{k}_{12} =
-i(\overleftarrow{\nabla}_1-\overleftarrow{\nabla}_2)/2\, .$
Here, the   right and left arrows indicate that the momentum operators act
on the right and  on the left,  respectively.  The  Skyrme parameters
$t_i$, $x_i$ and $\alpha$ for a fixed value of $W_0$ can
be expressed in terms of the quantities associated with the symmetric
nuclear matter as follows \cite{Chabanat97,Margueron02,Gomez92}.
\begin{eqnarray}
t_0&=&\frac{8}{\rho_{nm}}\left [\frac{\left (-{B/A}+\left (2m/m^*-3\right )\left (\hbar^2/10m\right )k_f^2\right )\left (\frac{1}{27}K_{nm}-\left (1-6m^*/5m\right )\left (\hbar^2/9m^*\right )
k_f^2\right )}{-{B/A}+\frac{1}{9}K_{nm}-\left (4m/3m^*-1\right )\left
(\hbar^2/10m\right )k_f^2}\right . \nonumber\\
&+&\left . \left (1-\frac{5m}{3m^*}\right )\frac{\hbar^2}{10m}k_f^2\right ],
\label{t0}
\end{eqnarray}
\begin{equation}
t_1=\frac{2}{3}\left [T_0 + T_s\right ],
\end{equation}
\begin{equation}
t_2=t_1+\frac{8}{3}\left[\left (
\frac{1}{4}t_0+\frac{1}{24}t_3\rho_{nm}^\alpha\right )
\frac{2m^*}{\hbar^2}\frac{k_f}{\pi^2}+G_0^\prime\right ]
\frac{\hbar^2}{m^*\rho_{nm}},
\label{t2}
\end{equation}
\begin{equation}
\label{t3}
t_3=\frac{16}{\rho_{nm}^{\alpha+1}}\frac{\left (-{B/A}+\left (2m/m^*-3\right )\left (\hbar^2/10m\right )k_f^2\right )^2}
{-{B/A}+\frac{1}{9}K_{nm}-\left (4m/3m^*-1\right )\left (\hbar^2/10m\right )k_f^2},
\end{equation}
\begin{equation}
x_0=\frac{4}{t_0\rho_{nm}}\left [\frac{\hbar^2}{6m}k_f^2 -\frac{1}{24}t_3
(x_3+\frac{1}{2})\rho_{nm}^{\alpha+1} + \frac{1}{24}\left (t_2\left
(4+5x_2\right )-3t_1x_1\right )\rho_{nm} k_f^2 -J \right]- \frac{1}{2},
\end{equation}
\begin{equation}
x_1=\frac{1}{t_1}\left [4\frac{\hbar^2 \kappa}{m\rho_{nm}}-t_2(2+x_2)\right ] -2,
\label{x1}
\end{equation}
\begin{equation}
x_2=\frac{1}{4t_2}\left [8T_0-3t_1-5t_2\right ],
\end{equation}
\begin{equation} 
x_3=-\frac{8}{\alpha t_3\rho_{nm}^{\alpha+1}}\left
[\frac{\hbar^2}{6m}k_f^2-\frac{1}{12}\left ((4+5 x_2)t_2-3t_1 x_1\right
)\rho_{nm} k_f^2-3 J +L\right ]-\frac{1}{2},
\label{x3}
\end{equation}
\begin{equation}
\alpha=\frac{{B/A}-\frac{1}{9}K_{nm}+\left (4m/3m^*-1\right )\left (\hbar^2/10m\right )k_f^2}
{-{B/A}+\left (2m/m^*-3\right )\left (\hbar^2/10m\right )k_f^2},
\label{alpha}
\end{equation}
where,

\begin{equation}
T_0=\frac{1}{8}\left (3t_1+(5+4 x_2)t_2\right )=
\frac{\hbar^2}{m\rho_{nm}}\left (\frac{m}{m^*}-1\right ),
\label{T0}
\end{equation}
\begin{equation}
T_s=\frac{1}{8}\left [9t_1-(5+4 x_2)t_2\right ],
\label{TS}
\end{equation}
and
\begin{equation}
k_{f}= \left (\frac {3\pi^2}{2}\rho_{nm} \right)^{1/3}.
\end{equation}
In Eqs. (\ref{t0})-(\ref{alpha}), the various quantities characterizing
the nuclear matter are the binding energy per nucleon $B/A$, isoscalar
effective mass $m^*/m$, nuclear matter incompressibility coefficient
$K_{nm}$, symmetry energy coefficient $J$,  the coefficient $L$ which
is directly related to the slope of the symmetry energy coefficient
($L=3\rho dJ/d\rho$), enhancement factor $\kappa$ and Landau parameter
$G_0^\prime$.  All these quantities are taken at the saturation density
$\rho_{nm}$.  It must be pointed out  that we include consistently all
terms in the energy density functional. In particular,  the expression
for the parameter $G_0^\prime$ used in the Eq. (\ref{t2}) is consistent
with the energy density functional obtained using Eq. (\ref{v12}). For
a more generalized Skyrme energy density functional, the expression
for  $G_0^\prime$ should be appropriately modified following 
Ref. \cite{Bender02}.  Once, $T_0$ is known, $T_s$ can be calculated
for a given value of the surface energy $E_s$ as \cite{Margueron02},
\begin{equation}
\label{Es}
E_s=8\pi r_0^2\int_0^{\rho_{nm}} d\rho
\left [\frac{\hbar^2}{36m}-\frac{5}{36}T_0\rho+\frac{1}{8}T_s\rho
-\frac{m^*}{\hbar^2}V_{\rm so}\rho^2\right ]^{1/2}
\left [B(\rho_{nm})/A-B(\rho)/A\right ]^{1/2}
\end{equation}
where, $B(\rho)/A$ is the binding  energy per nucleon given by,
\begin{equation}
\frac{B(\rho)}{A}=-\left [\frac{3\hbar^2}{10m^*}k_f^2+\frac{3}{8}t_0\rho+\frac{1}{16}t_3\rho^{\alpha+1} \right ]
\end{equation}
and,
\begin{eqnarray}
r_0=\left [ \frac{3}{4\pi\rho_{nm}}\right ]^{1/3}, \quad
V_{\rm so}=\frac{9}{16}{W_0}^2.
\end{eqnarray}
The manner in which Eqs. (\ref{t0}) - (\ref{alpha}) can be used
to evaluate the Skyrme parameters $t_i$, $x_i$  and $\alpha$ is as
follows. First, the parameters  $t_0$ and $\alpha$ can be obtained from 
Eqs. (\ref{t0}) and (\ref{alpha}),  respectively. Then, the parameter  $t_3$ can be calculated
using Eq. (\ref{t3}).  Next, $T_0$ and $T_s$ can be calculated using
Eqs. (\ref{T0}) and (\ref{Es}), respectively.  Once, the combinations
$T_0$ and $T_s$ of the Skyrme parameters  are known, one can calculate
the remaining parameters in the following sequence, $t_1$, $t_2$, $x_2$,
$x_1$, $x_3$ and $x_0$.

The stability criteria requires that \cite{Migdal67},
\begin{equation}
\label{ic}
{\cal X}_l > -(2l + 1),
\end{equation}
where, ${\cal X}_l$ stands for the Landau parameters $F_l$,
$F_l^\prime$, $G_l$ and $G_l^\prime$ for a given multipolarity $l$.
The Skyrme interaction only contains monopolar and dipolar contributions
to the particle-hole interaction so that all Landau parameters are
zero for $l > 1$. Thus, there are 12  different Landau parameters,
i.e.,  $F_l$, $F_l^\prime$, $G_l$ and $G_l^\prime$ ($l = 0, 1$) for the
symmetric nuclear matter and $F_l^{(n)}$, $G_l^{(n)}$ ($l= 0, 1$) for
the pure neutron matter. Each of these  Landau parameters must satisfy
the inequality condition given by Eq. (\ref{ic}). Using the  expressions
for the Landau parameters in terms of the Skyrme parameters,  given 
in Refs.  \cite{Margueron02},
one can obtain the  values of the Landau parameters at any density for a
given set of the Skyrme parameters. Thus, the critical density which is
nothing but the maximum density up to which all the inequality conditions
are met can be easily determined.

We have compiled the values of $\kappa$, $L$ and $G_0^\prime$ for
several parameterization of the Skyrme interaction presented in Refs.
[2,3,11-16]. We find that the values of $\kappa$, $L$ and $G_0^\prime$
vary over a wide ranges  $0 - 2$, $40 - 160$ MeV and $-0.15 - 1.0$,
respectively. This is due to the fact that the experimental data used in
the least-square procedure to fit the values of the Skyrme parameters
can not constrain well the values of these quantities.  These quantities
can be constrained by requiring a reasonable value for the critical
density.  In what follows, we shall refer to the set of standard values
for six quantities  $\rho_{nm}= 0.16 $fm$^{-3}$, ${B/A} = 16$ MeV, $K_{nm}
=230 $ MeV, $m^*/m =0.7$, $E_s =18$ MeV and $J =32 $MeV in short as  the
STD values for the nuclear matter input as used in Ref. \cite{Margueron02}.

We now study the  dependence  of  $\rho_{cr}$ on the $\rho_{nm}$,
$B/A $, $m^*/m$, $K_{nm}$, $E_s$ and $J$.  For a given set of values
for these quantities, we calculate the maximum value of $\rho_{cr}$
by varying $\kappa$, $L$ and $G_0^\prime$ at the $\rho_{nm}$ within
acceptable limits. We denote the maximum value of the critical density
by $\tilde\rho_{cr}$. It must be emphasized  here that the present
work differs from  the one performed in Ref. \cite{Margueron02} by the
fact that we constrain the values of $\kappa$, $L$ and $G_0^\prime$,
whereas, in Ref. \cite{Margueron02}, the value of $\tilde\rho_{cr}$
was calculated by varying the combinations $t_ix_i$ ($i=1, 2,$ and
3) of Skyrme parameters  with no  restrictions. We find that if the
space of the parameters $t_ix_i$ is not restricted, it may lead to an
unreasonable values of $\kappa$ and $L$.  As an example, for the STD
values of nuclear matter input, $\tilde\rho_{cr}$ becomes $3.5\rho_0$
for $\kappa = 1.0$, $L = 36$ MeV and $G_0^\prime = 0.20$ (at $\rho_0$).
The value of $G_0^\prime$  seems reasonable, but the value of  $\kappa =
1.0$ is a little too large \cite{Chabanat97,Keh91}. Further, we find
that for $\rho > \rho_0$ the value of  $L$ decreases with increasing
$\rho$ and it becomes negative for $\rho > 1.6\rho_0$, which makes the
interaction not favorable for the neutron star model.

In Fig. \ref{rhocr_std} we have displayed the results for
$\tilde\rho_{cr}$ obtained by varying  the various quantities associated
with the nuclear matter around their standard values. To calculate
$\tilde\rho_{cr}$ we allow for $0.25 \le \kappa \le 0.5 $, $0 \le L\le
100 $ MeV  and $0 \le G_0^\prime\le 0.5 $ at the $\rho_{nm}$. 
We further demand that  $L > 0$ at $3\rho_0$.  Fig. \ref{rhocr_std} shows
that $\tilde\rho_{cr}$ depends strongly on $m^*/m$ and $E_s$. Whereas,
$\tilde\rho_{cr}$ depends weakly on $\rho_{nm}$, ${B/A}$ and $K_{nm}$ and
it is almost independent of $J$.  These features for $\tilde\rho_{cr}$ are
qualitatively similar to the ones presented in Ref. \cite{Margueron02}.
However, due to the restrictions imposed on the values of $\kappa$,
$L$ and $G_0^\prime$, the values of  $\tilde\rho_{cr}$ becomes smaller
than that obtained in Ref. \cite{Margueron02}  by up to  $25\%$.
For $m^*/m = 0.6$ (0.7) and  keeping all the other nuclear matter
quantities equal to their standard values, we get $\tilde\rho_{cr} =
4.5\rho_0$ ($2.8\rho_0$) compared to $6\rho_0$ ($3.5\rho_0$)  obtained
in Ref. \cite{Margueron02}.

It may be instructive to present the values of $\kappa$, $L$ and
$G_0^\prime$  required to  obtained  $\tilde\rho_{cr}$ for a given set
of values for the nuclear matter input. We find that $\kappa$ lies in
the range of   $0.45 - 0.5$ for variations in the  nuclear matter input
by up to $\pm 15\%$ relative to their standard values. These values of
$\kappa$ clearly reflects the  fact that restricting $\kappa$ to take
values in the range of   $0.25 - 0.5$ delimits the $\tilde\rho_{cr}$ to a
lower value. In Figs. \ref{L-nm}a and \ref{L-nm}b we have plotted the
values of $L$ and $G_0^\prime$ (at $\rho_{nm}$), respectively, which are
needed to yield the $\tilde\rho_{cr}$.  We  see that $L$ varies from
$20$ to $ 60$ MeV for different values of the nuclear matter input.
For the STD values of nuclear matter input we find that $L = 47$ MeV
at $\rho = \rho_0$. This value is quite large compared to the values
of  $L = 35, 27$ and 16 MeV associated with the the Skz0, Skz1 and Skz2
interactions \cite{Margueron02}, respectively, which were obtained for
the same standard values of the nuclear matter input, but by varying the
combinations $t_1 x_1$ and $t_2 x_2$  with no restrictions and keeping
$t_3 x_3$ fixed to some arbitrary values.  We see from Fig. \ref{L-nm}b
that, except for $J$, the value of  $G_0^\prime$ (at $\rho_{nm}$) depends
strongly on the values of the various quantities associated with nuclear
matter. The dependence of $G_0^\prime$ on the surface energy coefficient
$E_s$ is the most pronounced one.  We note that $E_s$ is mainly determined
by the ground state properties of light nuclei. Thus, the center of mass
correction to the binding energy and charge radii must be appropriately
taken into account as they  are very important for light nuclei and may
affect the values obtained for  $E_s$.  We see from Fig. \ref{L-nm}b
that $G_0^\prime$ tends to vanish  rapidly with increasing $E_s$.

As pointed out earlier (see from Fig. \ref{rhocr_std}), the dependence of
$\tilde\rho_{cr}$ on $\rho_{nm}$,  $J$, and ${B/A}$ is quite weak.  Thus,
it may be sufficient to calculate $\tilde\rho_{cr}$ as a function of
$m^*/m$, $E_s$ and $K_{nm}$ only.  In Fig. \ref{efm_es_k230} we display
our results for the variation of $E_s$ versus $m^*/m$,  obtained for
fixed values of $\tilde\rho_{cr}$ with the remaining nuclear matter
quantities kept equal to their standard values. It can be seen from
Fig. \ref{efm_es_k230} that for a fixed value of $\tilde\rho_{cr}$,
$E_s$ decreases with the increase in $m^*/m$.  It is quite interesting to
note that for $E_s = 18 \pm 1$ MeV (as most of the Skyrme interactions
yield), $\tilde\rho_{cr} = 2\rho_0$ and $3\rho_0$ for $m^*/m = 0.72 -
0.85$ and $0.63 - 0.73$, respectively. For $\tilde\rho_{cr} = 4\rho_0$
one must have $m^*/m \sim 0.65$ for not too low  value of $E_s$. The
value of $m^*/m$ is also constrained by the  centroid energy of the
isoscalar giant quadrupole resonance \cite{Reinhard99_1} which favors
$m^*/m \ge 0.7$. Thus, for  reasonable values of $E_s$ and $m^*/m$,
one may obtain a Skyrme interaction with $\tilde\rho_{cr} = 2\rho_0$
to $3\rho_0$.

In summary, we have  used the  stability conditions of the Landau
parameters for the symmetric nuclear matter and pure neutron matter
to calculate the critical densities for the Skyrme type effective
nucleon-nucleon interactions.  We find that the critical density can
be maximized by appropriately adjusting the values of  the enhancement
factor $\kappa$, coefficient $L$ and the Landau parameter $G_0^\prime$
as these quantities are not well determined by the Skyrme parameters,
conventionally obtained by fitting the experimental data for the
ground state properties of finite nuclei. We exploit the fact that i)
The value of $\kappa$ should be in the range of $0.25  - 0.5$,   needed
to describe the TRK sum rule for the isovector giant dipole resonance
\cite{Chabanat97,Keh91}, ii) $L > 0$ for $ 0 \le \rho \le 3\rho_0$;
a condition necessary for a Skyrme interaction to be suitable for
studying the  properties of neutron star \cite{Stone03}.  and iii)
$G_0^\prime > 0$ in order to reproduce the energies  of the isovector
$M1$ and Gamow-Teller states \cite{Keh91}.  The maximum value of the
critical density so obtained is lower by up to $25\%$  compared to the
ones obtained without any such restrictions \cite{Margueron02}.  We show
that  the critical density obtained  for realistic  values of the surface
energy coefficient  ($E_s = 18 \pm 1$ MeV) and isoscalar effective mass
($m^*/m = 0.7 \pm 0.1$) lie in the range of $2\rho_0$ to $3\rho_0$.
Finally, we would like to  remark that the pairing correlations were not
included in the present work. It may be worthwhile to  investigate the
influence of the pairing correlations on the various quantities associated
with the nuclear matter which may alter the value of the critical density.

\bigskip
\begin{acknowledgments}

This work was supported in part by the US Department of Energy under Grant
No. DOE-FG03-93ER40773 and by the National Science Foundation under the
Grant No. PHY-0355200.
\end{acknowledgments}

\newpage
\begin{figure}[ht]
\caption{\label{rhocr_std} The dependence of critical density
$\tilde\rho_{cr}$ on the relative  variation of $\rho_{nm}$ (dotted),
${B/A}$  (dashed), $m^*/m$ (solid),  $K_{nm}$ (open circles),  $E_s$
(dashed-dot), and $J$ (dashed-filled squares) around their standard values.}

\caption{\label{L-nm} Variation of (a) the coefficient  $L $ and (b) the
Landau parameter $G_0^\prime$ as a function of the various quantities
associated with nuclear matter at $\rho_{nm}$.  The values of $L$ and
$G_0^\prime$ are determined by maximizing the critical density for a
given set of values for the nuclear matter quantities.}

\caption{\label{efm_es_k230} Variations of the  surface energy
coefficient  $E_s$ at $\rho_{nm}$ as a function of the effective mass
$m^*/m$   for fixed values of the critical density $\tilde\rho_{cr}
= 2\rho_0, 3\rho_0$ and $4\rho_0$ as labeled.  All the other nuclear
matter quantities are kept equals to their standard values.}

\end{figure}

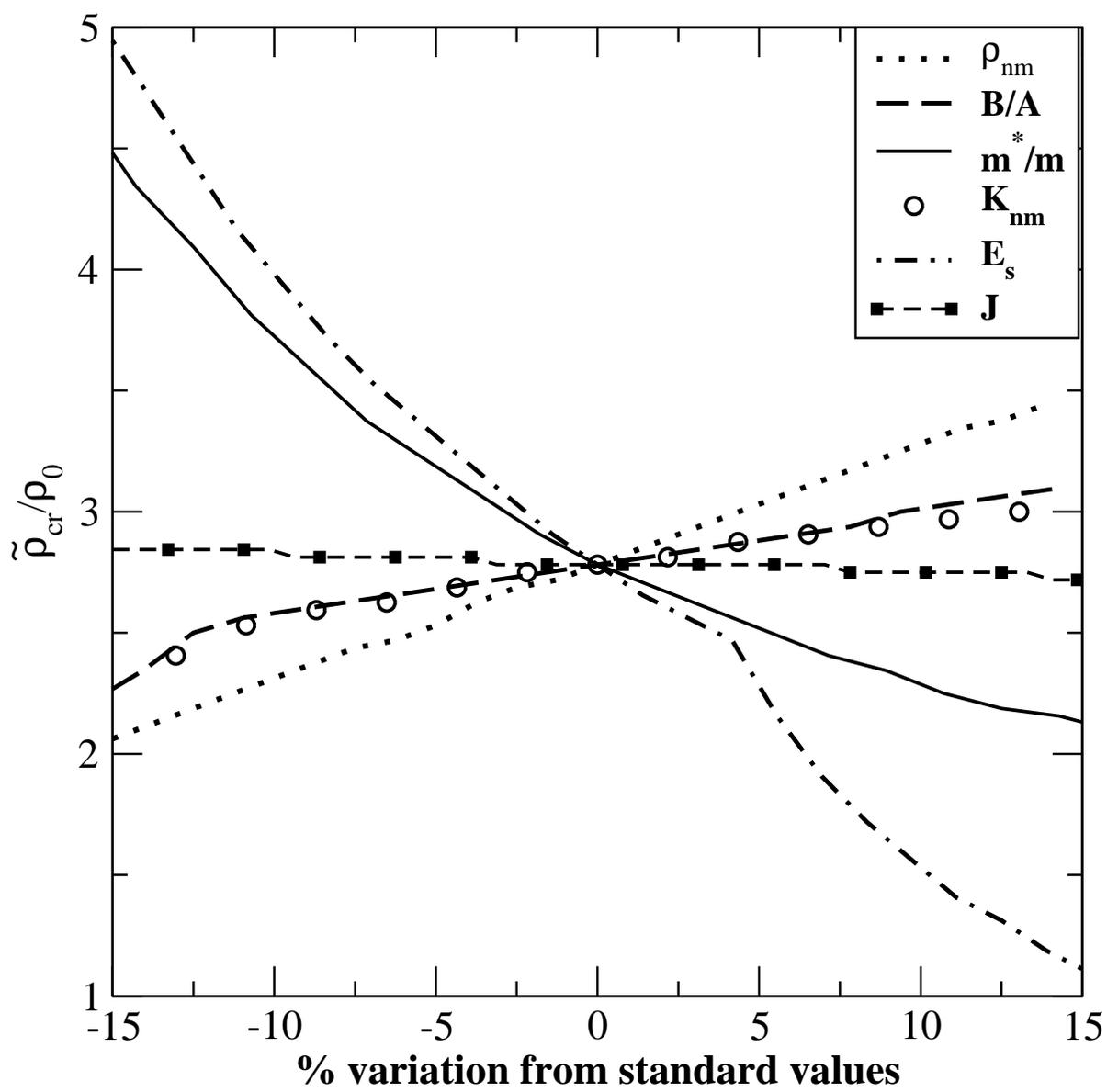
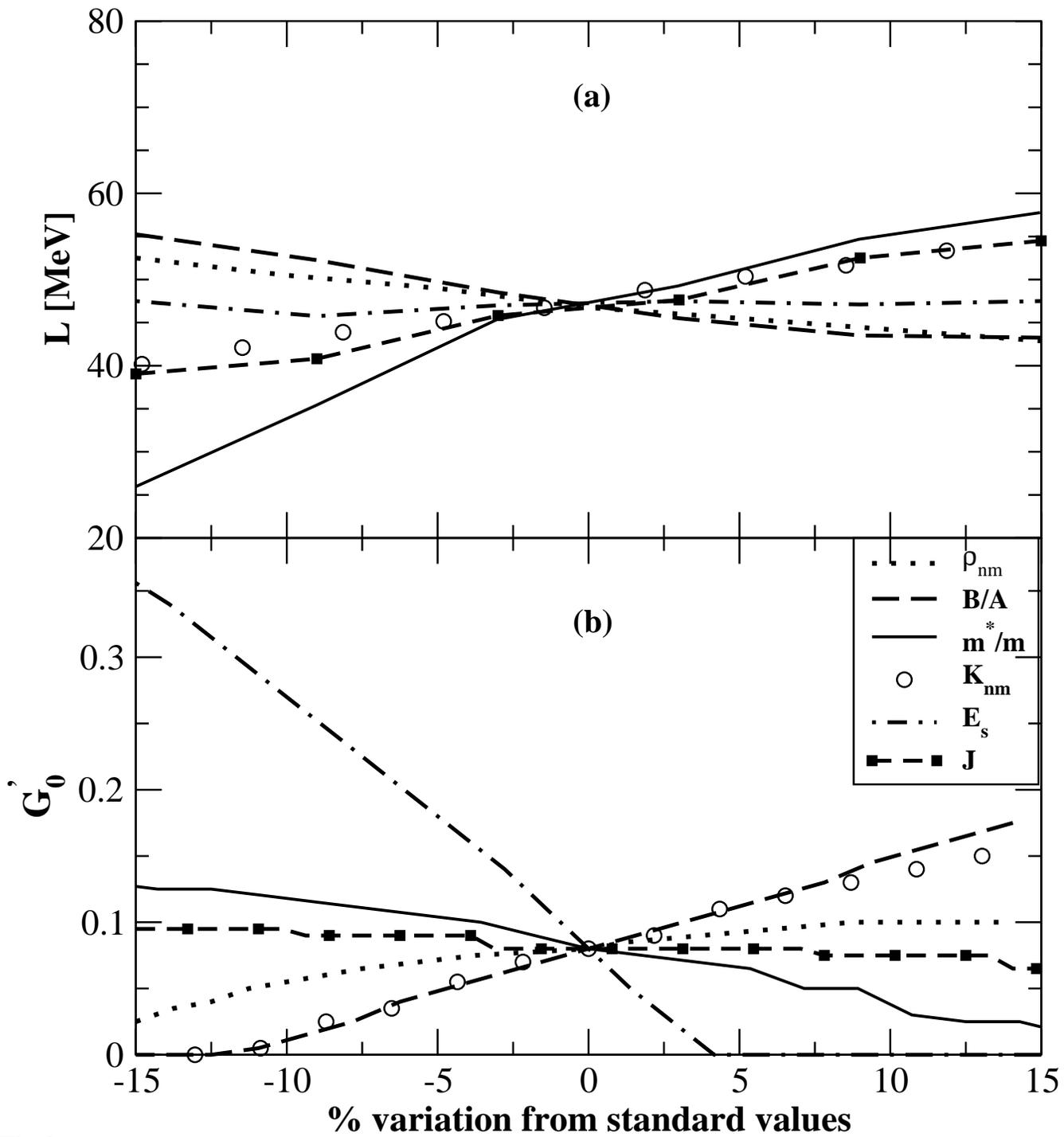
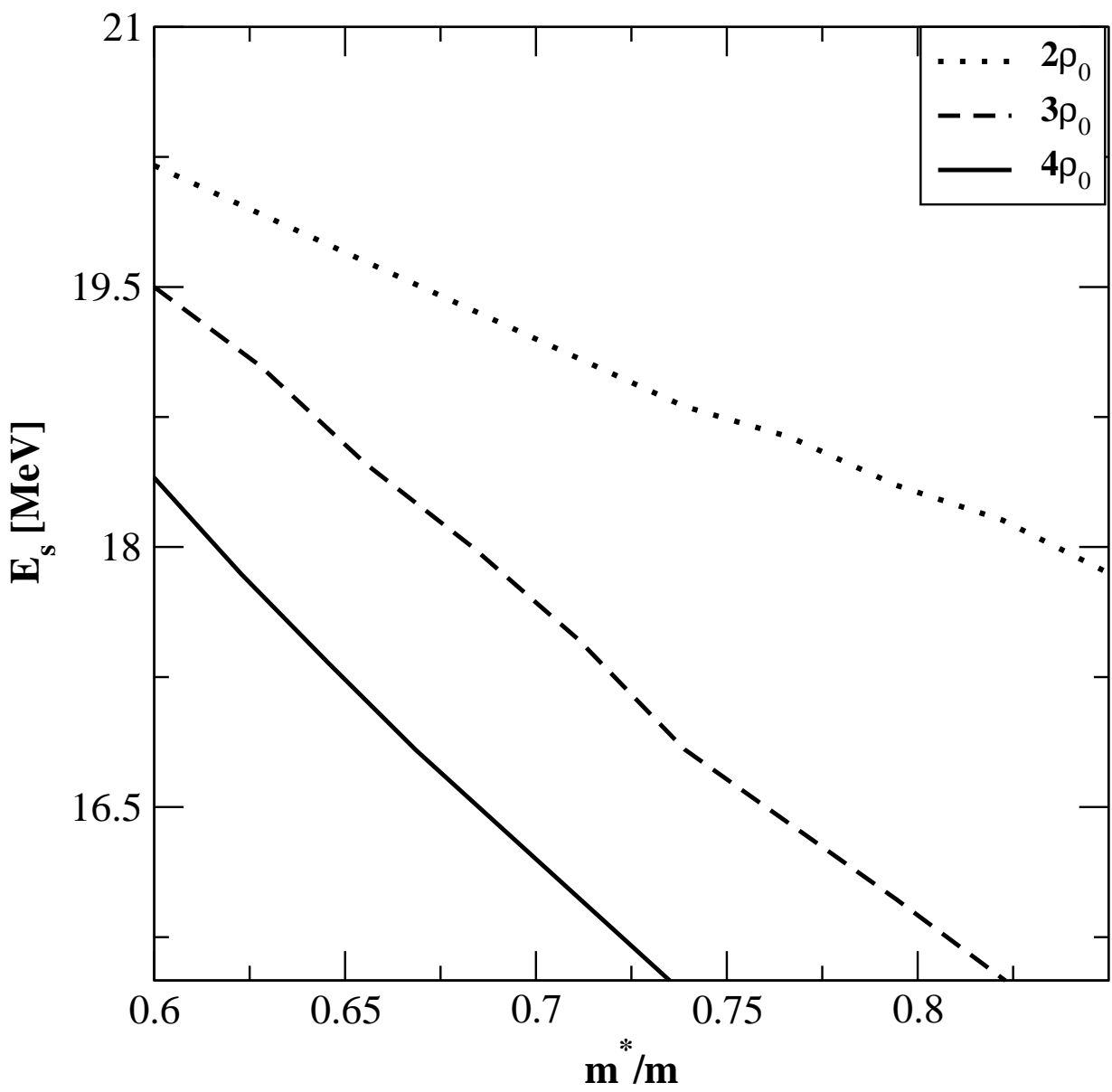
\end{document}